\documentclass[aps,prl,twocolumn,letterpaper,superscriptaddress, showpacs, amsmath, amssymb]{revtex4}

\usepackage{graphicx}
\usepackage{amsmath,amssymb}
\usepackage{color}

\begin{document}


\title{Subphotospheric Neutrinos from Gamma-Ray Bursts: The Role of Neutrons}

\author{Kohta Murase}
\affiliation{Hubble Fellow --- Institute for Advanced Study, Princeton, New Jersey 08540, USA}
\affiliation{Center for Cosmology and AstroParticle Physics; Department of Physics, The Ohio State University, Columbus, Ohio 43210, USA}

\author{Kazumi Kashiyama}
\author{Peter M\'esz\'aros}
\affiliation{Department of Astronomy and Astrophysics; Department of Physics; Center for Particle and Gravitational Astrophysics, Pennsylvania State University, University Park, Pennsylvania 16802, USA}

\date{submitted 17 January 2013; published 26 September 2013}

\begin{abstract}
Relativistic outflows with neutrons inevitably lead to inelastic collisions, and resulting subphotospheric $\gamma$ rays may explain prompt emission of $\gamma$-ray bursts.  In this model, hadronuclear, quasithermal neutrinos in the 10--100~GeV range should be generated, and they may even have a high-energy tail by neutron-proton converter or shock acceleration mechanisms.  We demonstrate the importance of dedicated searches with DeepCore+IceCube, though such analyses have not been performed.  Successful detections enable us to discriminate among prompt emission mechanisms, probe the jet composition, and see roles of relativistic neutrons as well as effects of cosmic-ray acceleration.  
\end{abstract}

\pacs{98.70.Rz, 95.85.Ry \vspace{-0.3cm}}

\maketitle


Gamma-ray bursts (GRBs) are the most luminous astrophysical phenomena with the isotropic $\gamma$-ray luminosity, $L_\gamma\sim{10}^{52}~{\rm erg}~{\rm s}^{-1}$.  Prompt $\gamma$ rays are observed in the MeV range, and their spectra can often be fitted by a smoothed broken power law (PL)~\cite{grbrev}. 
The emission is considered to be radiated from a relativistic jet with the Lorentz factor of $\Gamma\sim{10}^{3}$.  Observed light curves are highly variable down to $\sim1$~ms, suggesting unsteady outflows.  

Internal shocks are naturally expected for such unsteady jets, and the jet energy can be converted into radiation via shock dissipation.  In the classical scenario~\cite{rm94}, $\gamma$ rays are attributed to optically thin synchrotron emission from nonthermal electrons accelerated at internal shocks.  But there are troubles in explaining observational features such as the low-energy photon index, the high radiation efficiency and spectral correlations~\cite{prompt}.  

A promising alternative is the photospheric scenario, where prompt $\gamma$ rays are generated around or under the ``photosphere'' (where the Thomson optical depth $\tau_T$ is unity)~\cite{photosphere1}.  Since the emission is largely thermal, this scenario has advantages to explain the high radiation efficiency and stabilize the peak energy~\cite{photosphere2}.  Observations have also indicated a thermal-like component~\cite{fermigrb}.  In particular, subphotospheric dissipation may originate from inelastic nucleon-neutron collisions just beyond the decoupling radius~\cite{neutron1,neutron2}.  This ``inelastic collision model'' naturally predicts a broken PL $\gamma$-ray spectra via electromagnetic cascades and Coulomb heating~\cite{bel10}.  

The GRB prompt emission mechanism has been a long-standing, big mystery~\cite{grbrev}.  Different dissipation mechanisms are considered in the photospheric scenario, and optically thin models including the classic and magnetic reconnection scenarios are also viable (e.g., Ref.~\cite{magdis}).  Hence, discriminating among the various models is crucial, and neutrinos are powerful for this purpose since they can probe physical processes at subphotospheres ($\tau_T\gtrsim1$).    

In this work, we demonstrate the importance of dedicated searches for sub-TeV neutrinos.  Not only IceCube~\cite{icecube} but also its low-energy extension DeepCore~\cite{deepcore} are crucial for this purpose.  There are three key points. 
(1) Quasithermal neutrinos are inevitably produced via hadronuclear ($pp/pn/nn$) reactions, when inelastic collisions are responsible for the jet dissipation.  This is very different from classical and many magnetic scenarios, where neutrinos are mainly produced via the $p\gamma$ reaction between sufficiently high-energy cosmic rays (CRs) and photons~\cite{grbnu}. 
(2) Detecting sub-TeV neutrinos supports the photospheric scenario~\cite{mur08,wd09}, allowing us to reveal the prompt emission mechanism and probe the jet composition (e.g., baryon loading) and acceleration at subphotospheres. 
(3) In addition, we can study roles of relativistic neutrons on CR acceleration, including the neutron-proton-converter (NPC) acceleration mechanism~\cite{npc,kas+13} that may be relevant in low-luminosity~\cite{llgrb} and failed~\cite{mw01} GRBs.   

We hereafter use $Q_x \equiv Q/{10}^x$ in cgs units.  


{\it Inelastic collision model.---}
We consider the inelastic collision model, where dissipation is mainly caused via hadronuclear reactions at subphotospheres~\cite{bel10}.  Neutron-loaded jets are naturally expected in GRB engines including accretion disks and protoneutron stars~\cite{engine}.  In the baryonic fireball scenario, the jet Lorentz factor finally achieves $\Gamma_r\approx\eta$, which is the initial value of random internal energy per particle.  Initially, protons and neutrons are well coupled, but they are decoupled when the dynamical time is shorter than the elastic scattering time~\cite{neutron1}.  If the decoupling happens before coasting, neutrons form the slower flow with $\Gamma_s=\Gamma_n$.  Then, the faster flow with $\Gamma_r$ naturally overtakes the slower flow, leading to inelastic collisions.  Even if the coasting is earlier, inhomogeneity in the jet leads to internal collisions at $r\approx2\Gamma_s^2r_i$~\cite{rm94}, where $r_i$ is the jet basis.  

Considering an internal collision between outflows with $\Gamma_r$ and $\Gamma_s$, the Lorentz factor of the $\gamma$-ray emitting region is $\Gamma\approx\sqrt{\Gamma_r\Gamma_s/(\Gamma_s/\Gamma_r+\tau_{pn})}$ (when the faster flow collides with the neutron flow at $\tau_{pn}\leq1$)~\cite{bel10,px94}.  Here $\tau_{pn}\approx n_n\sigma_{pn}(r/\Gamma_n)$ is the optical depth for the $pn$ reaction, where $n_n\approx L_n/(4\pi\Gamma_n^2r^2m_nc^3)$ and $L_n$ is the neutron luminosity.  The $pp/pn$ cross section is $\sigma_{pn}\approx3\times{10}^{-26}~{\rm cm}^2$, and $\tau_{pn}=1$ corresponds to the dissipation radius of $r\simeq1.1\times{10}^{11}~{\rm cm}~(5L_n/L)L_{52}\Gamma_{n,2}^{-3}$.  Here $L$ is the kinetic luminosity of the interacting flow with $\Gamma$.  When the colliding flows completely merge, we have $\Gamma\approx\sqrt{\Gamma_r\Gamma_s}$.  

The kinetic energy of the faster flow may dissipate via inelastic collisions, as neutrons in the slower flow are swept.  Then, quasithermal nucleons with relativistic temperatures (with $\varepsilon_N^{\rm th}\approx\kappa_{p}\Gamma_{\rm rel}m_pc^2\simeq1.5~{\rm GeV}~\Gamma_{\rm rel,0.5}$ in the comoving frame of the interacting flow) are produced.  Here $\kappa_p\approx0.5$ is the nucleon inelasticity and $\Gamma_{\rm rel}\approx0.5(\Gamma/\Gamma_s+\Gamma_s/\Gamma)$ is the relative Lorentz factor between the interacting and slower flows.  Mesons and muons should be produced, which decay into $\gamma$ rays, electrons (positrons) and neutrinos~\cite{neutron1,neutron3}.  High-energy $\gamma$ rays cannot avoid the $\gamma\gamma$ process, and they induce electromagnetic cascades.  The cascades increase the number of pairs, so the Thomson optical depth is enhanced compared to $\tau_T$ for baryon-associated electrons.  When the pair density is determined by the balance between Coulomb heating by protons and inverse-Compton cooling, one obtains $\tau_T\approx23{(Y_\pm/0.2)}^{1/2}{(L/5 L_n)}^{1/2}{(\Gamma/5\Gamma_n)}^{-1}\tau_{pn}$~\cite{bel10}, where $Y_\pm$ is the pair yield.  In the subphotospheric dissipation, the $\gamma$-ray emission is still largely thermal via modification by Compton scatterings~\cite{photosphere1,photosphere2}.  The expected peak energy may be $\sim4~{\rm MeV}~\epsilon_\gamma L_{52}^{1/4}r_{i,7}^{-1/2}$ (where $\epsilon_\gamma$ is the radiation fraction), compatible with observations.  This thermal interpretation is consistent with time evolution observed in some GRBs~\cite{photosphere3}.  
Coulomb heating~\cite{bel10} or turbulence~\cite{slowheat} serves as slow heating, naturally leading to broken PL spectra, and a higher-energy component is formed by pair injections via the cascades.  

The high radiation efficiency is also naturally expected in the photosphere scenario~\cite{photosphere1,photosphere2}.  The energy carried by quasithermal nucleons is ${\mathcal E}_{\rm th}^{\rm iso}\sim0.5 {\mathcal E}_{N}^{\rm iso}$, where ${\mathcal E}_N^{\rm iso}$ is the kinetic energy that dissipates.  The significant fraction of the dissipated and trapped energy ($\sim 0.5 {\mathcal E}_{N}^{\rm iso}$) can be released as $\gamma$ rays.  Assuming that half of the energy is used for adiabatic expansion, we expect $\xi_N\equiv{\mathcal E}_{N}^{\rm iso}/{\mathcal E}_{\gamma}^{\rm iso}\sim4$.  The inelastic collisional model predicts $\xi_N\approx4\mbox{--}20$~\cite{bel10}.  


{\it Quasithermal neutrinos.---}
Sub-TeV neutrino production is the inevitable consequence of inelastic collisions.  Importantly, since neutrinos easily leave the flow, predictions for the hadronuclear neutrinos are insensitive to details of how to shape $\gamma$-ray spectra.  When the faster flow is decelerated by collisions with neutrons, the observed neutrino energy is typically $E_\nu^{\rm qt}\approx0.05\Gamma_rm_pc^2$.  Using the relative Lorentz factor between the fast and interacting flows, $\Gamma_{\rm rel}^\prime\approx0.5(\Gamma_r/\Gamma+\Gamma/\Gamma_r)\approx\Gamma_{\rm rel}\tau_{pn}$, we obtain 
\begin{equation}
E_\nu^{\rm qt} \approx0.1\Gamma\Gamma_{\rm rel}^\prime m_pc^2\simeq150~{\rm GeV}~\Gamma_{2.7}\Gamma_{\rm rel,0.5}^\prime,
\end{equation}
implying $\sim30\mbox{--}300$~GeV neutrinos for $\Gamma\sim{10}^2\mbox{--}{10}^3$.  A neutrino typically carries 1/4 of the pion energy but this energy fraction ranges from 0 to 0.43 in $\pi^\pm$ decay, and the high-energy tail is important for the detectability of neutrinos.  It is reasonable to take $\Gamma_{\rm rel}\sim{\rm a~few}$. 

{\it Nonthermal neutrinos.---}
When both the flows contain protons, internal shocks form and nonthermal neutrinos can be produced.  Note that neutrons can go through the faster flow when the neutron penetration depth $\sim{(\kappa_p \sigma_{pn} n_N)}^{-1}$ is longer than $r/\Gamma$.  A plausible possibility is the NPC acceleration mechanism~\cite{npc,kas+13}.  Hadronuclear reactions with incoming neutrons inevitably generate relativistic nucleons in the downstream, and protons are quickly isotropized by magnetic fields while experiencing the $pp$ reaction.  Then, as in the shock acceleration mechanism, a fraction of neutrons produced as a result of $n\rightarrow p\rightarrow n$ can go back into the upstream, overtaken by the shock front after the next conversion.  Quasithermal nucleons can be boosted by $2\kappa_p^2\Gamma_{\rm rel}^2$, so we obtain 
\begin{equation}
E_{\nu}^{\rm NPC}\approx0.05\Gamma(2\kappa_p^2\Gamma_{\rm rel}^2)\varepsilon_N^{\rm th}\simeq190~{\rm GeV}~\Gamma_{2.7} \Gamma_{\rm rel,0.5}^3.
\end{equation}
In principle, further boosts can be relevant when $\Gamma_{\rm rel}$ is large.  But using Eq.~(2) is typically enough due to other cooling processes~\cite{kas+13}.  Note that the NPC acceleration time must be comparable to the hadronuclear reaction time and the NPC acceleration mechanism is efficient only when $\tau_{pn}$ is not small.  Naive considerations lead to the efficiency $\epsilon_{\rm NPC}\approx g_{\rm NPC}(\Gamma_{\rm rel}^2/8){\rm min}[1,\tau_{pn}]$ and Monte Carlo simulations suggest $g_{\rm NPC}\sim0.03\mbox{--}0.3$~\cite{kas+13}.  

We also consider a possible PL component with the index $s$, assuming the proton acceleration time $t_{\rm acc}\approx2\pi\varepsilon_p/(eBc)$~\cite{grbnu}.  Its normalization is given by the efficiency $\epsilon_{\rm acc}$.  CR acceleration is possible especially at the reverse shock if $\tau_T\lesssim{\rm a~few}$, by, e.g., reducing the optical depth enhancement with smaller $Y_\pm$, though it is inefficient at radiation-mediated shocks~\cite{rms}.  
 

{\it Neutrino spectra.---}
We numerically calculate neutrino spectra with GEANT4~\cite{geant4}, following Ref.~\cite{mur08}.  
For the quasithermal component, we calculate hadronuclear reactions between cold nucleons with $\Gamma_{\rm rel}$ and incoming neutrons with $\Gamma_{\rm rel}$.  Quasithermal nucleons with $\varepsilon_N^{\rm th}$ are assumed to be isotropic and to lose their energies via various cooling processes.  Here we consider energy losses by Coulomb scattering, hadronuclear reactions, Bethe-Heitler process, photomeson production, synchrotron and inverse-Compton emission and adiabatic expansion.  At sufficiently high energies, decaying pions, kaons, and muons can lose their energies via hadronic processes, radiative cooling, and adiabatic expansion, which are treated by solving kinetic equations.  Practically, such cooling of mesons and muons are relevant only for a PL component.  There are four principal parameters, which are set to $\Gamma=600$, $\Gamma_{\rm rel}=3$, $\tau_{pn}=1$ and ${\mathcal E}_N^{\rm iso}=4{\mathcal E}_\gamma^{\rm iso}(\approx2{\mathcal E}_{\rm th}^{\rm iso})$.  The former two determine neutrino energies, while the latter two do the fluence normalization.  We also use a subparameter $L_n=2\times{10}^{51}~{\rm erg}~{\rm s}^{-1}$.  The comoving photon temperature is assumed to be $kT\approx530~{\rm eV}~L_{\gamma,52}^{1/4}r_{11}^{-1/2}\Gamma_{2.7}^{-1/2}$ and the magnetic energy fraction is set to $\epsilon_B=0.01$, though they are not critical for results on sub-TeV neutrinos.  

We also consider the NPC component that is approximated by a monoenergetic distribution with $0.5\Gamma_{\rm rel}^2\varepsilon_N^{\rm th}$ (in the comoving frame), and its normalization is set by $\epsilon_{\rm NPC}$ based on Ref.~\cite{kas+13}.  For the PL component, we determine the proton maximum energy $\varepsilon_p^{\rm max}$ by $t_{\rm acc}<{\rm min}[t_p,t_{\rm dyn}]$, where $t_p$ is the proton cooling time and $t_{\rm dyn}\approx r/(\Gamma c)$ is the dynamical time~\cite{mur08}.  Motivated by Ref.~\cite{ss11}, we adopt $s=2.1$ and $\epsilon_{\rm acc}=0.3$. 

The results for a high-luminosity GRB at $z=0.1$ are shown in Fig.~1.  As expected in Eq.~(1), quasithermal neutrinos have a peak at $\sim100$~GeV.  Importantly, their spectra are simply determined by hadronuclear reactions, insensitive to details of various cooling processes and how $\gamma$-ray spectra are shaped.  The NPC component is also shown with $\epsilon_{\rm NPC}=0.3$, which enhances a high-energy tail.  The possible PL component is prominent above TeV, and $p\gamma$ neutrinos can be dominant only at $\gtrsim0.1\mbox{--}1$~PeV.  

The atmospheric neutrino background (ANB)~\cite{icecubeatm} is also shown, assuming the angular window of max$[\Theta^2,\pi\theta_\nu^2]$, with $\Theta=5$~deg and the kinematic angle $\theta_\nu\approx1.5~{\rm deg}~\sqrt{{\rm TeV}/E_\nu}$.  Note that the neutrino mixing among the three flavors is properly taken into account.  

\begin{figure}[t]
\includegraphics[width=3.00in]{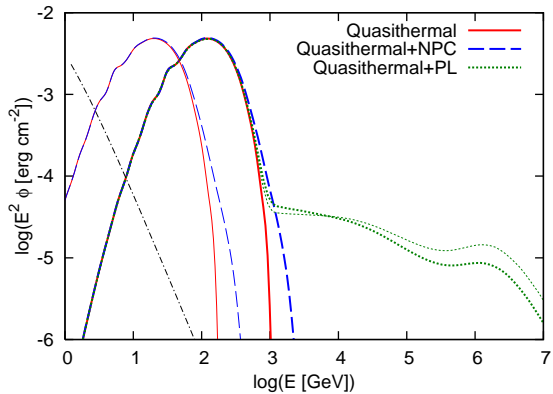}
\caption{The energy fluence of $\nu_\mu+\bar{\nu}_\mu$ from a high-luminosity GRB with ${\mathcal E}_\gamma^{\rm iso}={10}^{53.5}$~erg at $z=0.1$ (corresponding to $E_\gamma^2\phi_\gamma\sim{10}^{-2}~{\rm erg}~{\rm cm}^{-2}$).  The ANB in $30$~s is shown by the dot-dashed curve.  For solid and dashed curves, $\Gamma=600$ (thick) and $\Gamma=100$ (thin) are used.  For dotted curves, $s=2.1$ (thick) and $s=2.0$ (thin) are assumed.
}
\vspace{-1.\baselineskip}
\end{figure}

Detecting neutrinos from one GRB requires nearby bursts.  But most of these are much less energetic bursts like GRB 060218, which may originate from low $\Gamma$ jets~\cite{llgrb} or supernova shock breakouts~\cite{sbo}.  The results for a low-luminosity GRB at $D=10$~Mpc are shown in Fig.~2, with $\Gamma=30$, $\Gamma_{\rm rel}=5$, and a subparameter $L_n=2\times10^{46}~{\rm erg}~{\rm s}^{-1}$.  Quasithermal neutrinos are expected around 10~GeV, which also demonstrates lower $\Gamma$ cases.  The NPC component, which is prominent above $30\mbox{--}100$~GeV due to higher $\Gamma_{\rm rel}$, is shown with $\epsilon_{\rm NPC}=1$.  

\begin{figure}[t]
\includegraphics[width=3.00in]{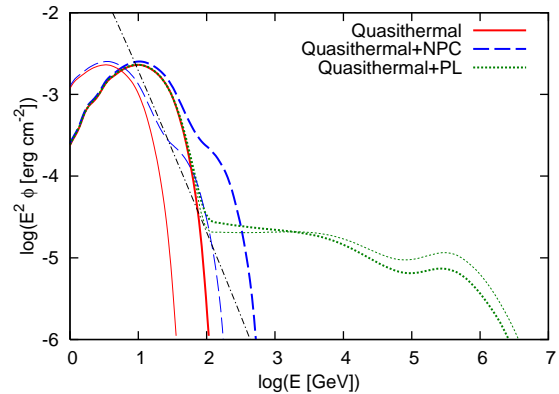}
\caption{The same as Fig.~1, but for a low-luminosity GRB with ${\mathcal E}_\gamma^{\rm iso}={10}^{50}$~erg at $D=10$~Mpc.  The ANB in $1000$~s is shown by the dot-dashed curve.  For solid and dashed curves, $\Gamma=30$ (thick) and $\Gamma=10$ (thin) are used. 
}
\vspace{-1.\baselineskip}
\end{figure}


{\it Neutrino detectability.---}
Since IceCube is not sensitive at $10\mbox{--}100$~GeV, including DeepCore is essential to see quasithermal neutrinos.  The neutrino effective area of DeepCore+IceCube at $10\mbox{--}100$~GeV is roughly $\approx{10}^{1.5}~{\rm cm}^2~{(E_\nu/100~\rm GeV)}^2$~\cite{deepcore}, so detections at $E_\nu$ require $E_\nu^2\phi_\nu\gtrsim 5\times{10}^{-3}~{\rm erg}~{\rm cm}^{-2}~{(E_\nu/100~\rm GeV)}^{-1}$.  For quasithermal neutrinos, we can roughly use $E_\nu\sim E_\nu^{\rm qt}\propto\Gamma$.  Only nearby and ``energetic'' GRBs can be seen, and a few events are detectable in the case shown in Fig.~1.  And the all-sky GRB rate within $z<0.1$ is only $\sim0.01\mbox{--}0.3~{\rm yr}^{-1}$~\cite{grbrate}, which is small and uncertain. 

Hence, we consider dedicated stacking analyses for GRBs detected by $\gamma$-ray satellites, although such analyses have been done around PeV energies for the classical scenario~\cite{icecubegrb}, but not at $\lesssim1$~TeV for the photospheric scenario.  To demonstrate how to search for subphotospheric neutrinos, we use the fluence distribution obtained by \textit{Fermi} GBM (see Fig.~7 in Ref.~\cite{gbm}).  About 300 bursts were observed in a year by GBM, \textit{Swift} and other spacecrafts~\cite{allgrb}, and GBM detected 400 long bursts in two years~\cite{gbm}.  We assume coincident detections of 3250 bursts in the Northern Hemisphere in 20 years.  To discover the signal, the signal-to-background for each burst should be sufficiently large.  From Fig.~1, the ANB at $\sim100$~GeV is $\sim{10}^{-6}~{\rm erg}~{\rm cm}^{-2}$, so the fluence threshold for stacking should be $\gtrsim{10}^{-6}~{\rm erg}~{\rm cm}^{-2}$.  Taking thresholds of $\lesssim{10}^{-6}~{\rm erg}~{\rm cm}^{-2}$ is not useful since the integrated fluence distribution is flat there, while using higher thresholds is not very essential since they are compensated by the smaller number of more energetic bursts.  

The expected number $N$ of detected $\nu_\mu+\bar{\nu}_\mu$ events is shown in Fig.~3, with the threshold of ${10}^{-5.5}~{\rm erg}~{\rm cm}^{-2}$.  The effective areas of DeepCore and IceCube are taken from Refs.~\cite{deepcore} and \cite{icecube}, respectively.  In this work, we adopt $\Gamma=600$ and $z=1$, and similar assumptions were also made in analyses in the classical scenario~\cite{icecubegrb}.  
How the neutrino fluence is normalized is crucial.  In the classical scenario, the normalization is given by a CR loading parameter or based on the observed ultrahigh-energy CR flux~\cite{grbnu}.  In the inelastic collision model, since subphotospheric $\gamma$ rays are responsible for the prompt emission, we can use the observed $\gamma$-ray fluence as ${\mathcal E}_N^{\rm iso}=\xi_N{\mathcal E}_{\gamma}^{\rm iso}\propto\xi_NE_\gamma^2\phi_\gamma$, adopting $\xi_N=4\mbox{--}20$.  This is analogous to the hadronic model for an extra GeV component~\cite{slowheat,asa+09}.  Note that we also consider the PL component, but it is as uncertain as ultrahigh-energy CR arguments.  In Figs.~3 and 4, we predict that a few events can be detected by analyzing $\sim1000\mbox{--}2000$ GRBs stacked in a decade.  Given the fluence at $E_\nu$, $N$ roughly decreases with $E_\nu$ in DeepCore, so $\sim3$ times as many GRBs are needed to find quasithermal neutrinos if all GRBs have $\Gamma\sim200$.  On the other hand, we can expect higher $\Gamma$ for energetic bursts, as suggested in LAT GRBs~\cite{fermigrb2}.  Quasithermal(+NPC) neutrinos lead to plateaus below $\sim100$~GeV due to their narrow distribution.  The possible PL component can enhance the detectability due to the large effective area of IceCube.  For a given $\epsilon_{\rm acc}$, the detectability changes by $\sim30$\% from $s=2.0$ to $s=2.2$~\cite{sbo} insensitively to $\Gamma$, but decreases as $\epsilon_{\rm acc}$.  

Muon neutrinos are mainly detected from muon tracks, whereas electron neutrinos are seen via showers (Fig.~4).   The ANB is more severe since the angular resolution is worse.  But better reconstruction techniques can improve the detectability significantly, e.g., if the low-energy extension of KM3Net could achieve $\sim 5$~deg~\cite{km3netcore}.

\begin{figure}[t]
\includegraphics[width=3.00in]{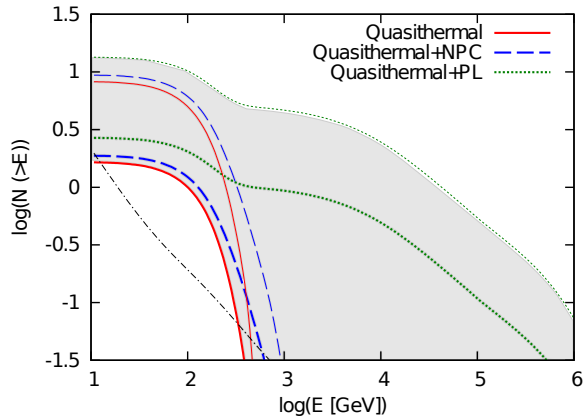}
\caption{The expected number of $\nu_\mu+\bar{\nu}_\mu$ events, which can be detected by coincident 20~yr observations with DeepCore+IceCube.  The dot-dashed curve is the ANB.}
\vspace{-1.\baselineskip}
\end{figure}

\begin{figure}[t]
\includegraphics[width=3.00in]{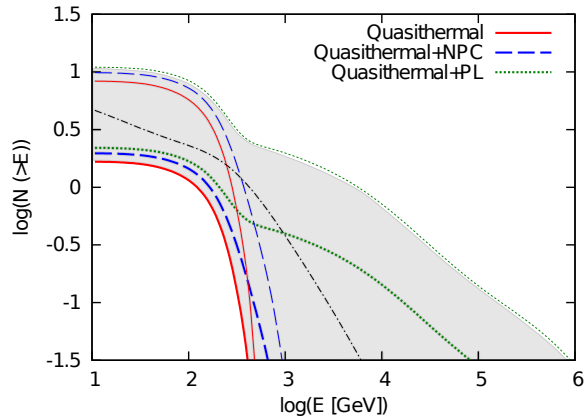}
\caption{The same as Fig.~3 but for $\nu_e+\bar{\nu}_e$ that can be observed via shower detections.  The angular resolution for the ANB is assumed to be 20~deg.}
\vspace{-1.\baselineskip}
\end{figure}


{\it Discussion and implications.---}
We showed that hadronuclear, quasithermal $10\mbox{--}100$~GeV neutrinos are inevitable when neutrons play a major role in generating prompt $\gamma$ rays.  For neutron-loaded jets, their signal is much more robust than more conventional nonthermal neutrinos that rely on uncertain CR acceleration mechanisms.  In the classical scenario, the $p\gamma$ reaction is dominant and its efficiency $f_{p \gamma}$ is sensitive to $r$ and $\Gamma$ that are uncertain~\cite{grbnu}.  The typical energy also depends on $\Gamma$ and $r$ (for sufficiently high $f_{p \gamma}$) as well as $L_\gamma$ and peak energy.  On the other hand, the inelastic collision model predicts $E_\nu\sim E_\nu^{\rm qt}$ and the $pn$ efficiency of $f_{pn}\approx\kappa_p\tau_{pn}\sim1$, and connections to observed CRs are unexpected due to strong cooling~\cite{mur08}.

Neutrons play various roles in jets at subphotospheres~\cite{neutron2}.  Quasithermal particles may naturally become seeds injected into CR acceleration processes, and detecting sub-TeV neutrinos provides insights into the NPC acceleration mechanism and possible CR acceleration.  In addition, neutrons may generate magnetic fields via $np$ conversions.  As neutrons go through the unmagnetized faster flow, they lead to proton beams or quasithermal protons with relativistic temperatures.  In particular, plasma anisotropies may lead to filamentation or Weibel instabilities, making the faster flow magnetized.  The magnetic fields are important for scattering of particles as well as synchrotron emission of electrons.

So far, dedicated searches have not been done and using only IceCube is insufficient.  This work strongly encourages stacking analyses with low-energy extensions of IceCube and KM3Net, and detections are possible in a decade with DeepCore-like detectors.  Coincident sub-TeV neutrinos can reveal the prompt emission mechanism that is a long-standing problem and provide information on the jet composition.  Successful detections may also give us clues to the jet acceleration via estimating $\Gamma$ from $E_\nu^{\rm qt}$.  Nearby low-luminosity and failed GRBs can also be interesting targets to reveal the GRB-supernova connection, especially with the NPC acceleration mechanism.

\medskip
We thank Imre Bartos, Doug Cowen, Boaz Katz and Szabolcs M\'arka for discussions.  We acknowledge support by the CCAPP workshop held in August 2012, Revealing Deaths of Massive Stars with GeV-TeV Neutrinos, where our preliminary results of this work were first presented.  This work is supported by NASA through Hubble Fellowship, Grant No. 51310.01 awarded by the STScI, which is operated by the Association of Universities for Research in Astronomy, Inc., for NASA, under Contract No. NAS 5-26555 (K. M.), JSPS fellowship for research abroad (K. K.) and NASA NNX13AH50G (P. M.).  When this work was completed, we became aware of the related, independent work~\cite{bar+13}, which was simultaneously submitted with this Letter.


\end{document}